\documentclass[5p,times]{elsarticle}

\makeatletter
\def\ps@pprintTitle{%
 \let\@evenfoot\@oddfoot}
\makeatother

\pdfoutput=1

\begin{document}

\begin{frontmatter}

\title{Universal atrial coordinates applied to visualisation, registration and construction of patient specific meshes}

\author[1]{Caroline H Roney}
\author[2,3]{Ali Pashaei}
\author[2]{Marianna Meo}
\author[2]{R\'emi  Dubois}
\author[4]{ Patrick M  Boyle}
\author[5]{ Natalia A   Trayanova}
\author[2]{Hubert   Cochet}
\author[1]{Steven A   Niederer}
\author[2,3]{Edward J   Vigmond}

\address[1]{School of Biomedical Engineering and Imaging Sciences, King's College London, United Kingdom}
\address[2]{LIRYC Electrophysiology and Heart Modeling Institute, Bordeaux Fondation, avenue du Haut-Lévèque, Pessac 33600, France}
\address[3]{IMB Bordeaux Institute of Mathematics, University of Bordeaux, 351 cours de la Libération, Talence 33405, France}
\address[4]{Department of Bioengineering, University of Washington, USA}
\address[5]{Department of Biomedical Engineering, Johns Hopkins University, USA}

\ead{edward.vigmond@u-bordeaux.fr}

\begin{abstract}
Integrating spatial information about atrial physiology and anatomy in a single patient from multimodal datasets, as well as generalizing these data across patients, requires a common coordinate system. In the atria, this is challenging due to the complexity and variability of the anatomy.
We aimed to develop and validate a universal atrial coordinate system for the following applications: combination and assessment of multimodal data; comparison of spatial data across patients; 2D visualization; and construction of patient specific geometries to test mechanistic hypotheses. 
Left and right atrial LGE-MRI data were segmented and meshed. Two coordinates were calculated for each atrium by solving Laplace's equation, with boundary conditions assigned using five landmark points. The coordinate system was used to map spatial information between atrial meshes, including atrial anatomic structures and fibre directions from a reference geometry.
Average error in point transfer from a source mesh to a destination mesh and back again was less than $6\%$ of the average mesh element edge length. 
Scalar mapping from electroanatomic to MRI geometries was compared to an affine registration technique (mean difference in bipolar voltage: $<10\%$ of voltage range). Patient specific meshes were constructed using UACs and phase singularity density maps from arrhythmia simulations were visualised in 2D.
We have developed a universal atrial coordinate system allowing automatic registration of imaging and electroanatomic mapping data, 2D visualisation, and patient specific model creation, using just five landmark points. 
\end{abstract}

\end{frontmatter}


\section{Introduction}
Characterisation of atrial electrophysiology and function requires comparison and integration of clinical measurements from different diagnostic modalities. These are measured across multiple intrapatient meshes to describe the individual, and across large interpatient datasets to describe populations. There is a large degree of variation in the shape, morphology and size of human atria, making standardised visualisation and analysis across large patient data sets challenging \cite{bisbal2013left}. 
Comparing measurements between geometries and creating patient specific meshes are technically challenging tasks, and may require the manual selection of a large number of landmark points, which is time consuming and subjective \cite{fahmy2007intracardiac}. 

The complexity of the atrial anatomy that makes registration challenging also hinders visualisation.  
Surface visualisation is more challenging for the atria due to their more complex shape. For left atrial visualisation, Karim et al. mapped the mitral valve (MV) annulus to the edges of a square and the pulmonary veins (PVs) to the centre using a technique that aimed to preserve areas and angles  \cite{karim2014surface};  Williams et al. further developed this method by creating a template mapping technique in which the MV annulus was mapped to a disk, the PVs to circles and the left atrial appendage (LAA) to an ellipse \cite{williams2017standardized}. Roney et al. developed a surface flattening technique to minimise distance distortion for two-dimensional visualisation of the left atrium (LA) \cite{roney2015technique} . However, this technique requires manual cutting of the mesh to determine the boundaries in two dimensions and it also has the limitation that it does not result in a standard coordinate system, and as such it is difficult to compare between geometries.

Registration algorithms enable comparison of data measured on different geometries, for example from different imaging or electroanatomic mapping systems. 
Atrial wall thickness \cite{bishop2015three}, activation \cite{cantwell2015techniques}, voltage \cite{rolf2014tailored} and late gadolinium enhancement \cite{cochet2018relationship} 
can all be measured across the atria in a single patient, and integrating these data through registration is key to understanding the mechanisms underlying atrial fibrillation. 
Machine learning algorithms, such as convolutional neural networks (CNNs), may be used to determine relationships between clinical measurables, but typically require the data to be organised in a standard spatial arrangement. 
As such, registration techniques are required but their application typically requires the manual selection of fiducial points and the meshes to have similar topologies.

Increasingly, integrated datasets are combined with physical and physiological constraints encoded in biophysical models. These human atrial computer models may include regional heterogeneity, for example electrophysiological and tissue conductance variations between the LA and RA body, PVs, appendages, pectinate muscles (PM) and Bachmann's bundle (BB), and also fibre direction. As such, the construction of detailed computational models requires the identification of different tissue regions and assignment of fibre directions. Previous studies have manually identified tissue types \cite{labarthe2012semi} or used mapping techniques \cite{krueger2013towards}. Typically, fibre directions are assigned using a rule based approach or mapped from an atlas using landmark points \cite{krueger2011modeling,fastl2016personalized,morgan2016slow}. 
Construction of large patient cohorts requires an automated registration and model generation platform. 

Here we develop a universal atrial coordinate (UAC) mapping system for application to the atria, motivated by our universal ventricular coordinate (UVC) system \cite{bayer2018universal}.
The UVC system consists of four coordinates, with simple boundary points at the apex and base. For the atria, we use just two coordinates for visualisation purposes, and the boundary conditions are defined along the valves, the roof, and the lateral and septal boundaries, which are visually evident landmarks for the atrial anatomy.  
We motivate the choice of coordinates, including the boundary conditions, and detail the methodology used for mapping both scalar and vector data, in order to visualise atrial data in two dimensions, to calculate correlations between measures, and to construct patient specific bilayer models with both electrophysiological heterogeneity and fibre direction. 
We show that the construction of the UAC system requires the manual selection of just five points. 
We further demonstrate that UACs may be used to generate patient specific bilayer meshes from segmented MRI datasets. We compare the technique to other scalar and vector registration techniques, and finally demonstrate an application of UACs for 2D visualisation.

\section{Methods}
\begin{figure}[!t]
\centering
\includegraphics[width=0.45\textwidth]{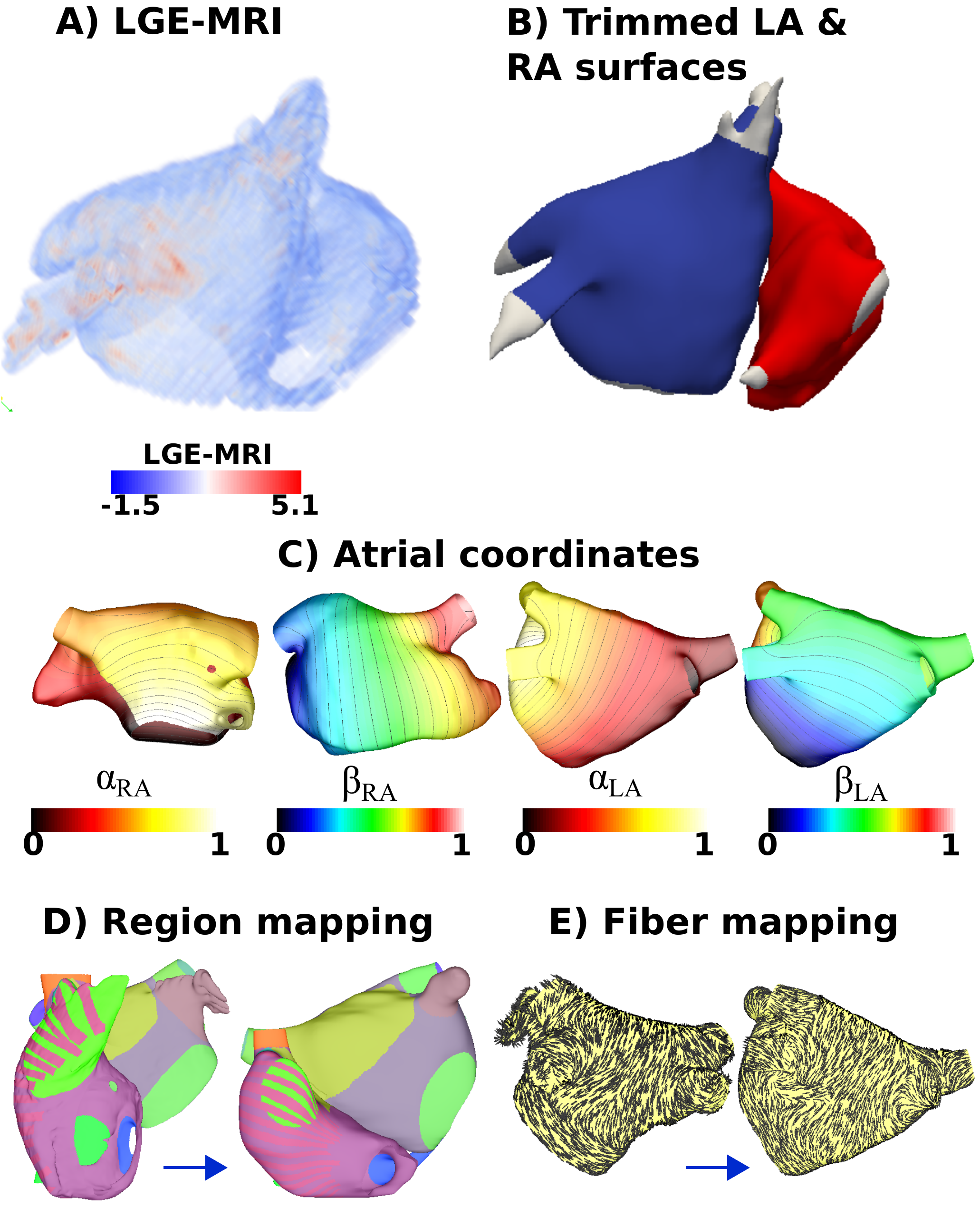}
\caption{\textbf{Schematic:} (A) The LA and RA were segmented from LGE-MRI data using Music software. (B) The endocardial segmentations were then meshed to create closed surface triangulations, and trimmed at the valves and veins (LA shown in blue; RA shown in red). (C) For each of the LA and RA surfaces, two atrial coordinates were defined.
For the RA, these were a lateral-septal TV coordinate ($\alpha_{RA}$, shown in septal-lateral view), and an IVC-SVC coordinate ($\beta_{RA}$, shown in lateral-septal view). 
For the LA, these were a septal-lateral coordinate ($\alpha_{LA}$, shown in posteroanterior view), and a posterior-anterior coordinate ($\beta_{LA}$, shown in posteroanterior view). 
(D) These coordinates were used to map atrial structures from the original reference mesh to the target patient specific mesh. (E) Vector fibre data were also mapped between geometries.}
\label{fig:sch}

\end{figure}

Our methodology maps a surface mesh of an atrium (left or right) to the unit square, allowing mapping of atlases, registration of geometries, and visualisation. 
We describe here the data types used (Section~\ref{sec:DM}), pre-processing employed (Section~\ref{sec:seg}-~\ref{sec:regions}) to generate the atrial meshes, followed by the point selection, and boundary conditions required for the calculation of the universal atrial coordinates (Sections~\ref{sec:lap}). We explain how to use these coordinates to map scalar (Section~\ref{sec:scalar}) and vector (Section~\ref{sec:vector}) data between geometries, which allows the co-registration of clinical data. Finally we describe an application of the UAC system to construct patient specific bilayer bi-atrial meshes, with repolarisation heterogeneity, atrial anatomical structures and fibre direction (Section~\ref{sec:patientspecific}). 
The schematic in Fig~\ref{fig:sch} outlines the methodology used for constructing patient specific bilayer geometries using the UAC system. 

\subsection{Data modalities} \label{sec:DM}
We used late-gadolinium enhancement (LGE) MRI data to generate patient specific geometries, as well as electroanatomic mapping data from the Carto system (Biosense Webster). 
This study complies with the Declaration of Helsinki and was approved by the Institutional Ethics Committee at the University of Bordeaux. All patients gave written informed consent.
Late-gadolinium enhancement MRI data were recorded at a resolution of 0.625mm x 0.625mm x 2.5mm, as previously described \cite{roney2016modelling}. The CARTO3 electroanatomic mapping system generated surface meshes for the LA with an average edge length of 1.9 mm. The methodology presented here is general and applicable to any atrial surface data.

\subsection{Segmentation and mesh modifications}\label{sec:seg}
Atrial geometry meshes were constructed using the following steps. 
Left and right atrial LGE-MRI data were segmented using MUSIC software (Electrophysiology and Heart Modeling Institute, University of Bordeaux, Bordeaux France, and Inria, Sophia Antipolis, France, http://med.inria.fr) (see Fig~\ref{fig:sch}A). 
Closed surface triangulations of these segmentations were generated using the IRTK mcubes algorithm \cite{schnabel2001generic}.
The triangulation for the LA was then cut at the MV and the four PVs were trimmed; correspondingly, the RA was cut at the tricuspid valve (TV), with the superior vena cava (SVC), the inferior vena cava (IVC), and coronary sinus (CS) trimmed using Paraview software (Kitware, Clifton Park, NY, USA) (see Fig~\ref{fig:sch}B). 
The fossa ovalis (FO) was identified from the segmentation as the area of intersection of the LA and RA meshes, and a circular area of each mesh corresponding to the overlapping region was removed, as for the original bilayer mesh \cite{labarthe2014bilayer}. 
The meshes were then remeshed using mmgtools meshing software (https://github.com/MmgTools/mmg), with parameters chosen to produce meshes with an average edge length of 0.34 mm, to match the resolution of the previously published bilayer model \cite{labarthe2014bilayer}. 

\subsection{Atrial region assignment}\label{sec:regions}
The following atrial regions were assigned to the model: the PVs, the SVC, the IVC, and the left and right atrial appendages (LAA and RAA). These structures are visually evident in the constructed LA and RA meshes, and as such may be labelled using a variety of methods; further details are given in the supplementary methods. These region labels were used in the UAC calculation to define the boundaries between the superior veins and the LA body, as well as between the junctions of the SVC and IVC with the RA body.

\subsection{Calculating the UACs}\label{sec:lap}
\subsubsection{Overview}
UACs were calculated by solving Laplace's equation on the LA or RA mesh with Dirichlet boundary conditions of zero and one applied along two sets of boundary nodes. 
Each of the LA and RA were parametrised using two coordinates selected to be as close to orthogonal as possible everywhere. 
For the LA, these coordinates were a septal to lateral coordinate, $\alpha_{LA}$, and a coordinate from the posterior MV to the anterior MV,  $\beta_{LA}$. 
Correspondingly for the RA, $\alpha_{RA}$ was from the lateral TV to the septal TV, and $\beta_{RA}$ passes from the IVC to the SVC. 
These directions were chosen as they result in the same LA and RA orientation that is used clinically for 2D visualisation of biatrial basket catheter data \cite{narayan2012computational}. 
The Laplace solves were performed using the CARP simulator \cite{vigmond2003computational}.  
The following sections describe the choice of boundary nodes used for each of the coordinates, and a rescaling required to define the final coordinates. We chose to define the LA and RA coordinates to be as similar to each other as possible. As such, we define equivalent roof boundaries using the superior veins only for the LA, and the venae cavae for the RA. 

\subsubsection{User input: point selection for the UACs}
\begin{figure}[!t]
\centering
\includegraphics[width=0.45\textwidth]{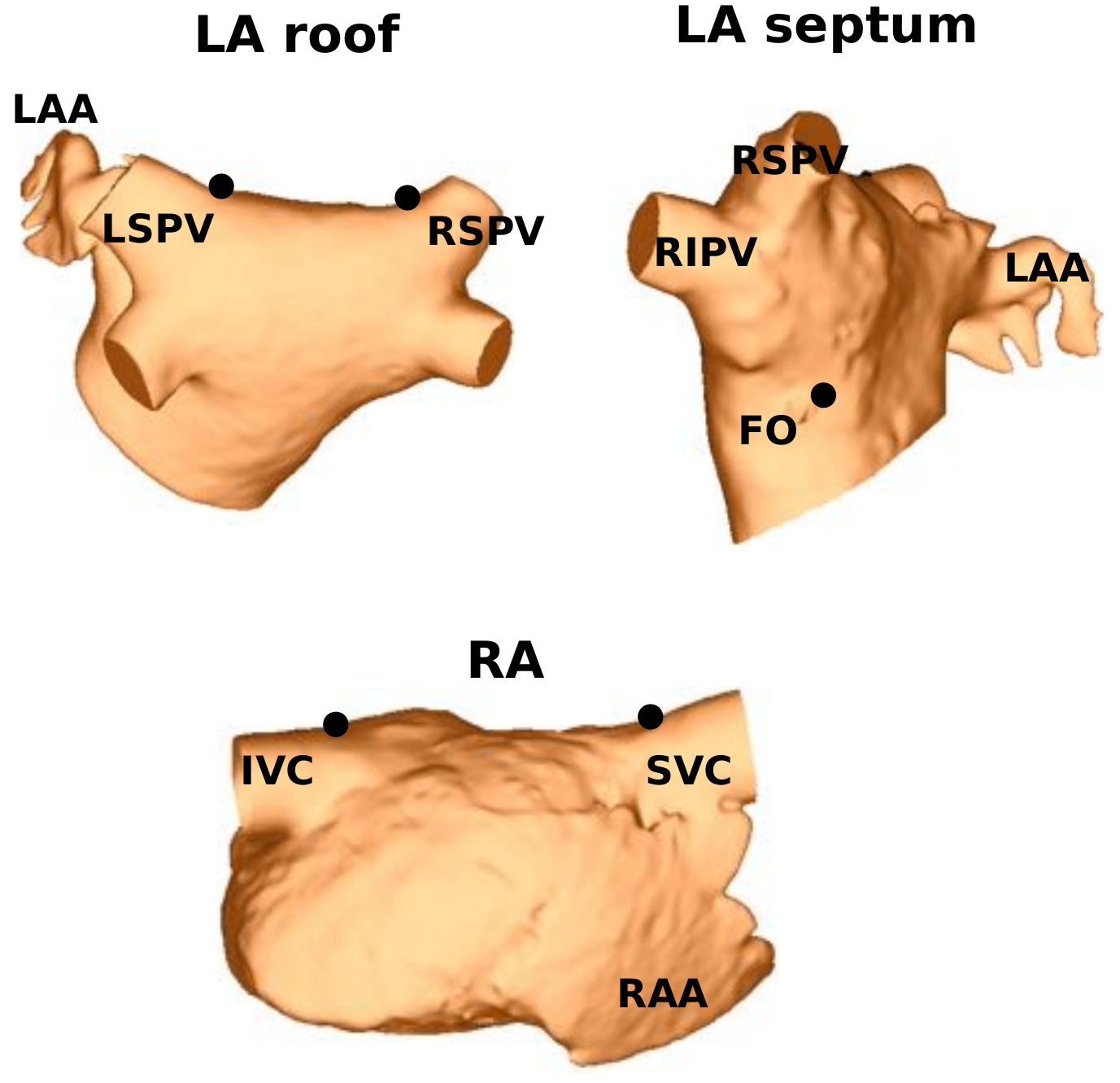}
\caption{\textbf{Selection of five landmark points:} Two points were chosen on the LA roof; one on the LA septum; and two on the RA lateral-septal boundary. 
The LA roof points were chosen at the junction of the LA body with the LSPV and RSPV, at the highest posterior location for the LA.
The LA septal point was chosen to be just anterior of the FO location.
The RA lateral-septal points were chosen in a similar way to the LA roof points; that is at the junction of the RA body with the IVC and SVC, at the highest lateral location for the RA. 
}
\label{fig:UACpoints}

\end{figure}

Calculation of the UACs requires the manual selection of three points for the LA and two points for the RA. For the LA, the geometry was initially rotated to be in the posteroanterior view, and a point chosen at the junction of the right superior PV (RSPV) with the LA body, and at the junction of the left superior PV (LSPV) with the LA body, at the boundary between the posterior and anterior walls (see Fig~\ref{fig:UACpoints}). These two points were used to construct a roof line to define how the geometry is divided into posterior and anterior components. 
A third point, determining the septal boundary, was chosen on the LA septum, at the anterior edge of the FO location (see Fig~\ref{fig:UACpoints}). 

For the RA, two points were chosen to construct a roof line, to divide the geometry into lateral and septal components. The RA geometry was rotated to be in the lateral view, and one point was chosen at the SVC-RA body junction and the other chosen at the IVC-RA body junction, both at the boundary between the lateral and septal walls (see Fig~\ref{fig:UACpoints}). The sensitivity of the mapping to these user inputs was investigated. 

\subsubsection{LA septal-lateral coordinate boundary nodes}\label{sec:LScoord}
The boundaries of the LA septal-lateral coordinate $\alpha_{LA}$ were constructed between the MV and the junctions of each of the superior veins with the LA body. 
To achieve this, first the MV ring nodes were identified as those in the largest connected edge list, and the RSPV-LA and LSPV-LA junctions were identified as nodes that are in elements from both regions (see Fig~\ref{fig:LapNodes} A). 
The septal boundary of the LA septal-lateral coordinate, $\alpha_{LA}$, consisted of the shortest geodesic path between the RSPV-LA junction nodes and the MV ring nodes that went through the user-defined FO point (see Fig~\ref{fig:LapNodes} B). 
Geodesic paths were calculated using a marching cubes algorithm \cite{peyre2009toolbox}. 
The lateral boundary was the shortest geodesic path between the MV ring nodes and the LSPV-LA junction nodes that was posterior of the LAA (see Fig~\ref{fig:LapNodes} B). 
To ensure that this path was posterior of the LAA, the path had to result in the LAA being on the anterior wall in the UAC system, which was automated by including a criterion that the left inferior PV (LIPV) and LAA were assigned to opposite walls (see Fig~\ref{fig:LapNodes} B).

\subsubsection{LA posterior-anterior coordinate boundary nodes}\label{sec:PAcoord}
\begin{figure}[!t]
\centering
\includegraphics[width=0.45\textwidth]{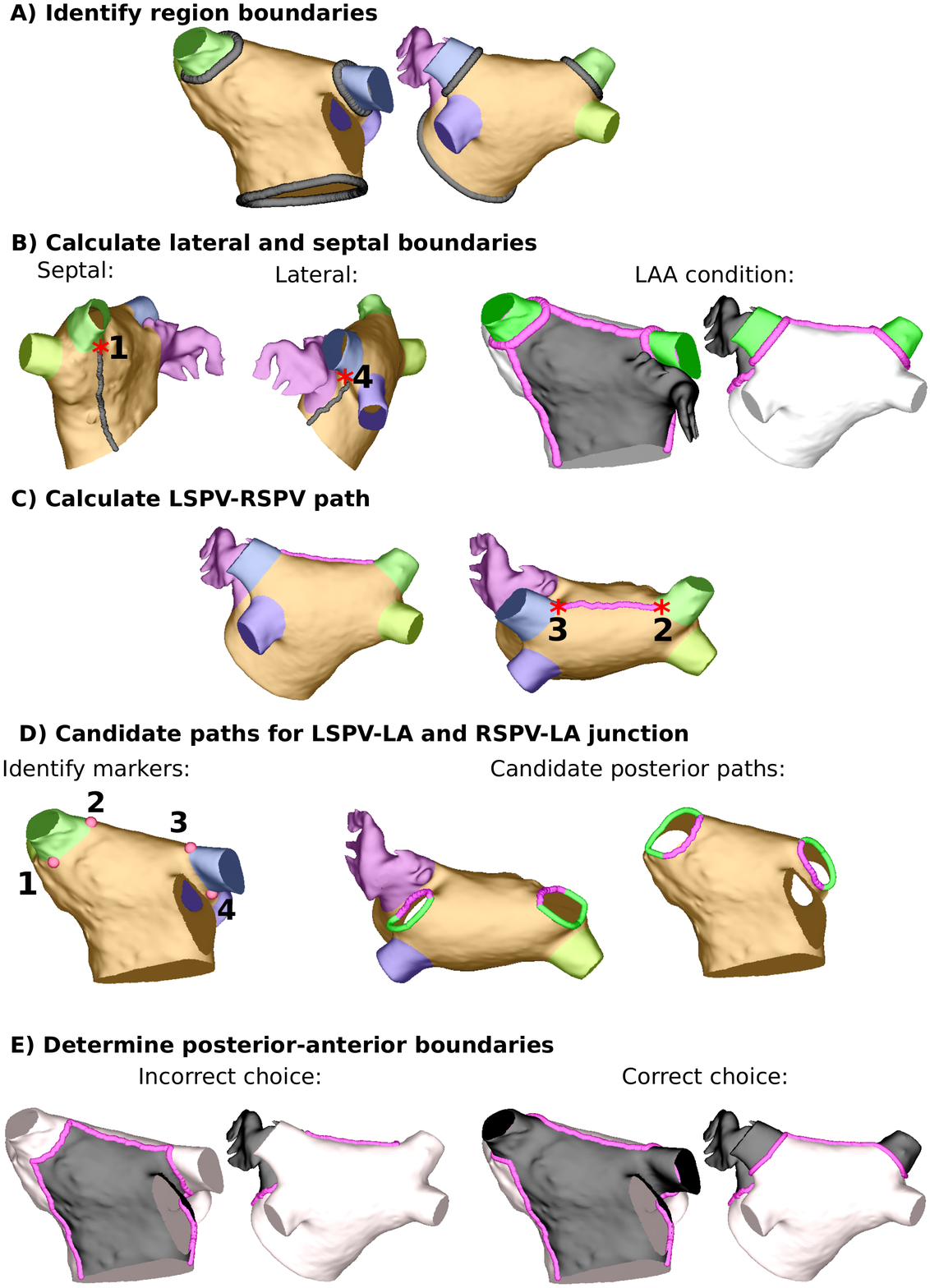}
\caption{\textbf{Calculation of Dirichlet boundary nodes used for Laplace solves.} (A) Identify region boundaries at the RSPV-LA and LSPV-LA junctions and MV (black lines). (B) Calculate the lateral and septal paths, ensuring the LAA is on the anterior wall (shown in black, with the posterior wall in white and superior veins in green). The intersections of these paths with the RSPV-LA and LSPV-LA junctions are marked (by 1 and 4, respectively).
(C) Calculate the geodesic path between the LSPV and RSPV user-defined markers (purple line). The intersections of this paths with the RSPV-LA and LSPV-LA junctions are marked (by 2 and 3, respectively).
(D) Use the four points (intersection of lateral, septal and roof boundaries with LSPV-LA and RSPV-LA junctions; points 1-4) to divide the superior vein openings into anterior and posterior segments (purple and green, respectively).
(E) Determine the anterior-posterior boundary choice (purple line) such that the superior veins are assigned as anterior (black region).} 
\label{fig:LapNodes}

\end{figure}

Calculation of the LA posterior-anterior coordinate ($\beta_{LA}$) required solving the Laplace equation with a set of boundary nodes, followed by a rescaling to separate the posterior and anterior components of the geometry. 
The first boundary was the MV ring, which was set to zero. 
The second boundary was the roof boundary nodes, which were set to one. This boundary consisted of the  geodesic path between the user-defined points at the posterior-anterior boundary of the RSPV-LA and LSPV-LA junctions (the LSPV-RSPV path, see Fig~\ref{fig:LapNodes} C), together with the posterior components of the junctions of the superior veins with the LA body.
Using only the posterior component of the junctions of the superior veins with the LA body ensures that the superior veins are well-defined in the UAC system.

To determine the posterior component of the superior PV-LA junctions, four markers were automatically identified (see Fig~\ref{fig:LapNodes}D) as follows: 1) where the septal boundary meets the RSPV-LA junction, 2) where the RSPV-LA junction meets the LSPV-RSPV path, 3) where the LSPV-RSPV path meets the LSPV-LA junction, and 4) where the LSPV-LA junction meets the lateral boundary. 
Each pair of points (1$\&$2 and 3$\&$4) divided the vein openings into two segments, one posterior and one anterior (see Fig~\ref{fig:LapNodes} D), but it was unknown which segment was which.
The posterior portion was then automatically selected as the path that mapped the superior veins to the anterior wall in the UAC system (see Fig~\ref{fig:LapNodes} E), which was determined by whether the LAA, LSPV and RSPV were path connected.

The aforementioned use of markers at the path boundaries ensures that the set of nodes consisting of the lateral, septal and roof nodes used to separate the posterior and anterior walls form a continuous path. 

\subsubsection{RA coordinate boundary nodes}
The RA lateral-septal TV coordinate ($\alpha_{RA}$) was calculated in a similar way to the LA posterior-anterior MV coordinate ($\beta_{LA}$, see Section~\ref{sec:PAcoord}). 
Specifically, the first boundary for the Laplace solve was the TV ring, which was set to zero. 
The second boundary, which was set to one, consisted of the geodesic path between the user-defined points at the lateral-septal boundary of the SVC-RA and IVC-RA junctions, together with the lateral components of the junctions of the SVC and IVC with the RA body. The lateral components of the SVC-RA and IVC-RA junctions were determined using the same technique as for $\beta_{LA}$. 

Correspondingly, the RA IVC-SVC coordinate ($\beta_{RA}$) was calculated in a similar way to the LA septal-lateral coordinate ($\alpha_{LA}$, see Section~\ref{sec:LScoord}). 
Specifically, the RA IVC boundary consisted of the shortest geodesic path between the IVC-RA junction nodes and the TV ring nodes. Similarly, the RA SVC boundary was the shortest path between the SVC-RA junction nodes and the TV ring nodes. The RA IVC boundary was chosen such that the CS was on the septal wall in the UAC system,  which was automated by including a criterion that the RAA and CS were assigned to opposite components of the geometry.

\subsubsection{Solving the Laplace equation}

Fig~\ref{fig:LP} B shows the Laplace solution for the IVC-SVC coordinate ($\beta_{RA}$), ranging between 0 on the IVC boundary and 1 on the SVC boundary. Similarly, the LA septal-lateral coordinate ($\alpha_{LA}$) ranges between 0 on the septal boundary (RSPV-MV path) and 1 on the lateral boundary (LSPV-MV path). 

Solving the Laplace equation for the posterior-anterior MV LA boundary nodes (to calculate $\beta_{LA}$), or the lateral-septal TV RA boundary nodes (to calculate $\alpha_{RA}$), 
resulted in a solution, $\psi$, ranging between 0 at the valve ring nodes and 1 at the LA roof or RA lateral-septal boundary (see Fig~\ref{fig:LP} A). 
This was then used to calculate a coordinate between 0 and 0.5 on the posterior LA wall (or equivalently the lateral RA wall), and between 0.5 and 1 on the anterior LA wall (or septal RA wall), as follows. The mesh was automatically separated into posterior LA (or lateral RA) and anterior LA (or septal RA) segments using the previously defined nodes. 
For the LA, these nodes consisted of the lateral and septal boundaries used for $\alpha_{LA}$, together with the roof nodes used for $\beta_{LA}$. For the RA, these nodes consisted of the SVC and IVC paths used for $\beta_{RA}$, together with the lateral-septal boundary used for $\alpha_{RA}$ (see Fig~\ref{fig:LP} C). 
Posteriorly $\beta_{LA}$ was then calculated as $\beta_{LA}= 0.5 \cdot \psi$, and anteriorly as $\beta_{LA}=1- 0.5 \cdot \psi$. 
Equivalently, laterally $\alpha_{RA}$ was then calculated as $\alpha_{RA}= 0.5 \cdot \psi$, and septally as $\alpha_{RA}=1- 0.5 \cdot \psi$ (see Fig~\ref{fig:LP} C). 

To ensure appropriate coordinate assignment for elements along the septal and lateral LA boundaries (or IVC and SVC RA boundaries), these boundary nodes (corresponding to $\alpha_{LA}$=0, $\alpha_{LA}$=1, $\beta_{RA}$=0, and $\beta_{RA}$=1) were duplicated and elements renumbered to the posterior or anterior choice of LA node (lateral or septal choice of RA node), as appropriate.

\subsubsection{Coordinate rescaling}\label{sec:rescale}
\begin{figure}[!t]
\centering
\includegraphics[width=0.45\textwidth]{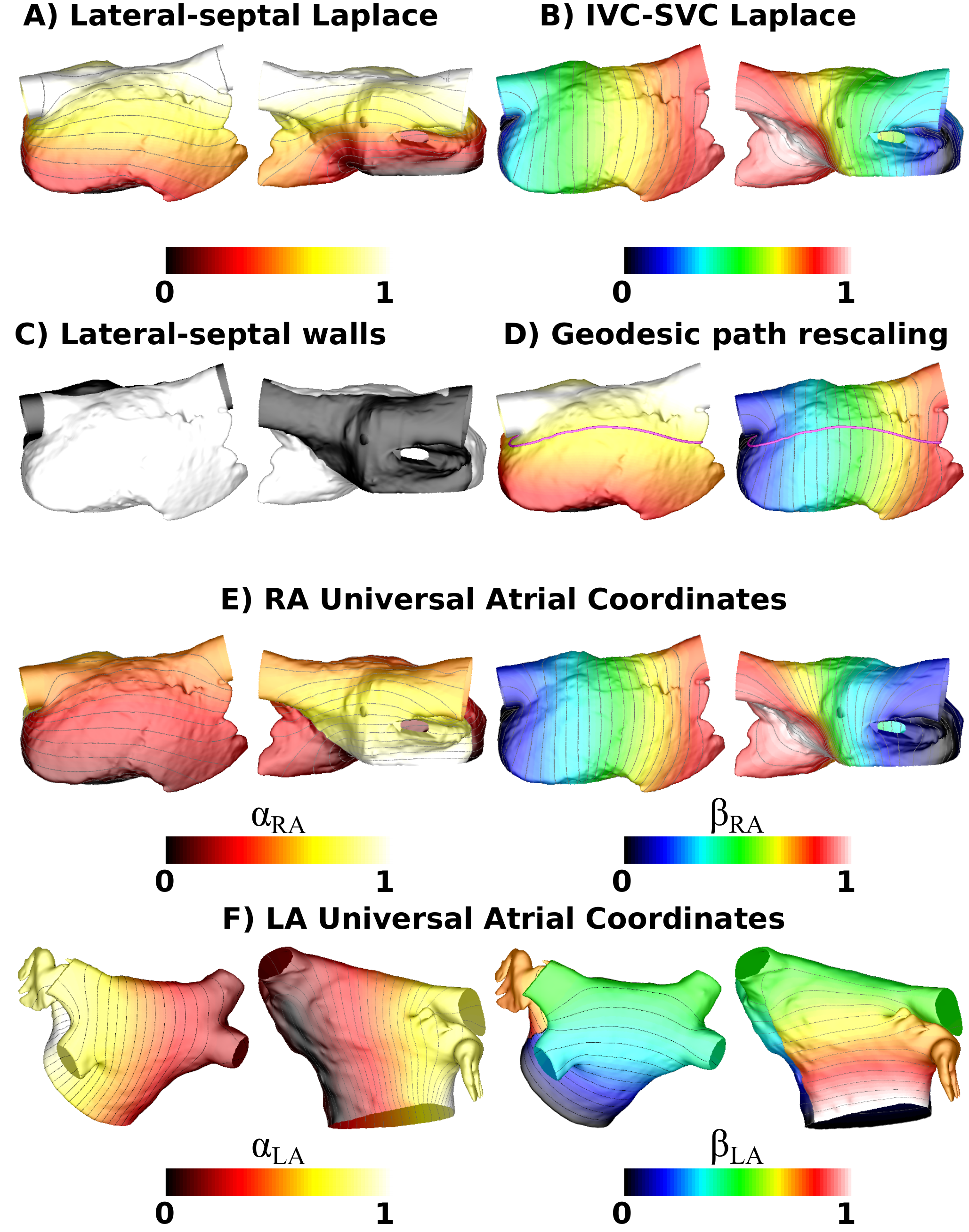}
\caption{\textbf{Laplace calculations and UACs.} 
(A) The Laplace field generated by solving for boundaries of 0 at the TV, and 1 along the lateral-septal boundary. 
(B) The Laplace field generated by solving for boundaries of 0 at the IVC path, and 1 along the SVC path. 
(C) Mesh partitioned into lateral and septal regions. 
(D) The lateral portion of an isoline (value 0.7) for the Laplacian solve in (A) was used to rescale the coordinate in (B). 
(E) UACs for the RA. 
(F) UACs for the LA.
Isolines are shown at 0.04 increments (except for A which is 0.08 spacing).} 
\label{fig:LP}

\end{figure}

The septal-lateral coordinate for the LA ($\alpha_{LA}$) and the IVC-SVC coordinate for the RA ($\beta_{RA}$) were rescaled to account for differences in atrial morphology between patients.
This was achieved for the LA by choosing a path spanning the posterior wall by automatically selecting nodes along a geodesic of constant posterior-anterior coordinate ($\beta_{LA}$). 
Equivalently, an RA path spanning the lateral wall was selected by automatically selecting nodes along a geodesic of constant lateral-septal TV coordinate ($\alpha_{RA}$) (see Fig~\ref{fig:LP} D).
Tracing along this geodesic yielded a mapping between the Laplace solution and the normalised geodesic distance. The mapping was then applied to the entire Laplace solution to arrive at the septal-lateral LA coordinate ($\alpha_{LA}$) and IVC-SVC RA coordinate ($\beta_{RA}$).
The final UACs are shown in Fig~\ref{fig:LP} E and F.

\subsection{Scalar value mapping}\label{sec:scalar}
To map scalar data between meshes, 
both the reference and the target geometry in $(x, y, z)$ were expressed in terms of the universal coordinates $(\alpha, \beta)$. For each element in the reference geometry, a change of basis matrix $M$ from $(x,y,z)$ to $(\alpha, \beta)$ was calculated and used to map the mid-point of the element in $(x,y,z)$ to $(\alpha, \beta)$. Similarly, mid-points in the target geometry were mapped to universal coordinate space. 

To determine a scalar field value for an element in the target geometry, we first identified the element of the reference geometry in universal coordinates that contains the mid-point of the target element. This mid-point of the target element was then expressed in barycentric coordinates of the reference geometry element. The scalar value was then assigned as a distance weighted average of the vertex values. For computational efficiency, only the five closest elements to the target mid-point were checked (assessed using their mid-points in universal coordinate space), and in the case that none of these elements contained the target point, the scalar value was assigned as a regular distance weighting of the three closest nodes. 
This use of barycentric interpolation was necessary for transferring data between meshes of different resolutions.

\subsection{Vector data mapping} \label{sec:vector}

To map vector data such as cardiac fibre directions between meshes, we followed the methodology used for the UVC system \cite{bayer2018universal}. 
Specifically, we constructed a local orthonormal basis at each point in the source mesh based on UAC, we then interpolated the vector field to this point, and finally projected the desired vector to this basis.
We then constructed a local orthonormal basis for each point in the destination mesh, 
and projected the vector direction in UACs from the source geometry to Cartesian space.

\subsection{Constructing patient specific bilayer meshes using scalar and vector mapping} \label{sec:patientspecific}
An application of the UAC system is constructing patient specific meshes, which may be used for simulations of cardiac electrophysiology, with regional heterogeneity and fibre direction. 
This construction utilises the previously published bilayer model \cite{labarthe2014bilayer} as an atlas model of atrial anatomy including the location of atrial structures and atrial fibres.
To construct bilayer meshes, first UACs are used to define several of the atrial structures and fibre directions; second, BB and the CS are incorporated using a sequence of rules; and finally the endocardial and epicardial layers are coupled using discrete resistance line connections. 

\subsubsection{Region and fibre vector assignment}
Both scalar and vector mapping were used to generate a bilayer mesh with endocardial and epicardial layers. 
The PVs, LAA, RAA, SVC and IVC regions were assigned using the techniques listed in the Supplementary material, and then scalar mapping using UACs was used to assign the CT, PM, sinoatrial node (SAN), line of block and sections of BB (following Section~\ref{sec:scalar}). 
These structures were encoded in the original LA and RA mesh by assigning labels to the elements in the original mesh that are closest to each of the elements in these structures. As such, the original mesh UACs may correspond to multiple structures. 
Elements in the LA and RA mesh of the new bilayer model were assigned atrial structure labels depending on the original mesh label, according to the closest element mid-point in UACs. Once these labels were assigned, additional endocardial right atrial structures (CT, PM, line of block, SAN) were added to the mesh by duplicating elements labelled as these structures and projecting their locations $0.1$mm endocardially, so that they appear on the endocardial but not the epicardial layer. 
The projection distance and arrangement of the atrial structures followed the original bilayer model \cite{labarthe2014bilayer}. 
The LA endocardial mesh was duplicated and projected $0.1$mm epicardially to generate two surface meshes to construct a bilayer model.  

Vector mapping of both the endocardial and epicardial LA fibres was performed using the LA UACs to assign these fibres to the target atrial geometry. Fibre directions were first assigned to the RA and RAA using UAC vector mapping, and then this was repeated on sections of the mesh representing the CT, PM and line of block regions to assign fibres to the corresponding structures. Vectors were normalised. 

BB and the CS require additional steps for their construction because they are not fully determined by the UAC system. Details for their inclusion in the model, and the use of discrete resistance line connections between atrial surfaces are described in the supplementary material.

 \subsubsection{Cylindrical fibres}
 In the instance that the universal coordinate system does not have complete coverage of an atrial structure, specific rules can be used to assign fibre directions. 
 For example, for cylindrical structures (such as the PVs, IVC and SVC) a possible choice is to assign 
 longitudinal fibre directions as a function of normalised distance along the cylinder. Fibre directions were modelled as parallel to the $z$ cylinder axis at the distal rim, and circumferential at the LA-PV junction, varying linearly within this $90^{\circ}$ range along the length of the cylinder.

 \section{Results}
 
\subsection{Point transfer}
To evaluate the accuracy of point transfer using UACs, we mapped the mesh vertices from one geometry to another and back again, and then computed the distance error of the mapped points from the original vertices. Specifically, following our UVC paper \cite{bayer2018universal}, we mapped $X^s  \rightarrow x^D  \rightarrow X^D  \rightarrow x^s$, where $x$ represents points in Cartesian space, $X$ represents points in UAC space, and the superscripts refer to the source (s) or destination (D) mesh. We mapped from both the LA and RA of the original bilayer model \cite{labarthe2014bilayer} to three different atrial geometries, and the average errors are given in Table~\ref{tab:pointtransfer}. 
Average mapping error for the LA was less than 2 $\mu$m and for the RA was less than 20 $\mu$m (average mesh edge length was 340 $\mu$m).

\begin{table}[!ht]
\centering
\caption{
{\bf Average error in $\mu$m for mapping from an atlas geometry to either the atlas or three different atrial geometries and back again.}}
\begin{tabular}{|l|l|l|l|l|l|}
\hline
\multicolumn{1}{|l|}{\bf}  & \multicolumn{1}{|l|}{\bf Atlas} & \multicolumn{1}{|l|}{\bf P1} & \multicolumn{1}{|l|}{\bf P2} & \multicolumn{1}{|l|}{\bf P3}\\ \hline
LA      
& $0.05$   
&$0.34$ & $2.12$  
& $0.16$    \\ \hline
RA      
&$10.48$   
& $5.93$ & $19.17$   
& $9.80$    \\ \hline
\end{tabular}
\label{tab:pointtransfer}
\end{table}

 \subsection{Scalar mapping and comparison to registration techniques}
\begin{figure}[!t]
\centering
\includegraphics[width=0.45\textwidth]{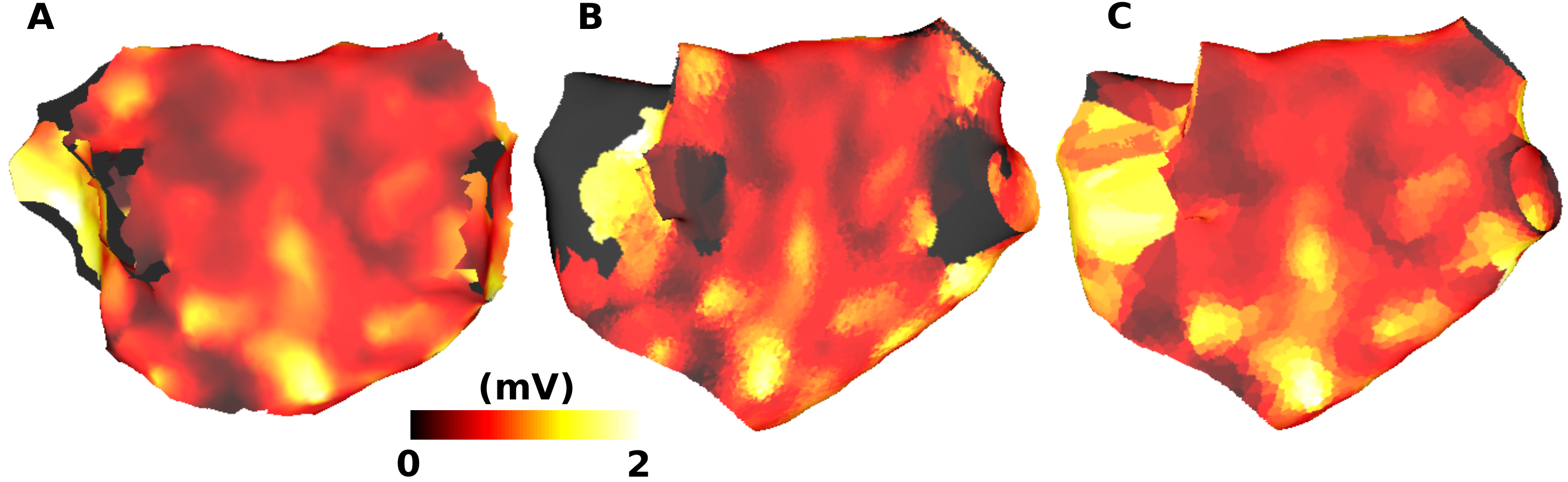}
\caption{\textbf{Mapping bipolar peak-to-peak voltage:} (A) Peak-to-peak voltage measured using the Carto system. (B) Peak-to-peak voltage mapped from (A) using UACs to a mesh created from LGE-MRI for the same patient. (C) Peak-to-peak voltage mapped from (A) using an affine registration technique.}
\label{fig:Carto}

\end{figure}

 One application of the UAC system is for mapping scalar data between different atrial geometries. 
 Here we compare UAC mapping to a standard registration technique for transferring electroanatomic voltage data to an MRI geometry, with the motivation of correlating electrical and structural data. 
 Specifically, the UAC system was used to transfer bipolar peak-to-peak voltage data measured using the CARTO3 electroanatomic mapping system to a mesh generated from MRI data for the same patient; shown in Fig~\ref{fig:Carto}. The peak-to-peak voltage maps are visually similar; the mapping using UACs was then compared to a mapping using standard registration techniques. 
 This registration was performed using the Medical Image Registration ToolKit (MIRTK, https://github.com/BioMedIA/MIRTK), using the \emph{register} function to perform an affine registration to the imaging data, followed by using the \emph{transform-points} function to apply this transformation to the surface vertices. 
 The average absolute point-wise difference in scalar values between the UAC (Fig~\ref{fig:Carto} B) and affine registrations (Fig~\ref{fig:Carto} C) is $0.26 \pm 0.26$mV.

 \subsection{Patient specific geometries}
\begin{figure}[!t]
\centering
\includegraphics[width=0.45\textwidth]{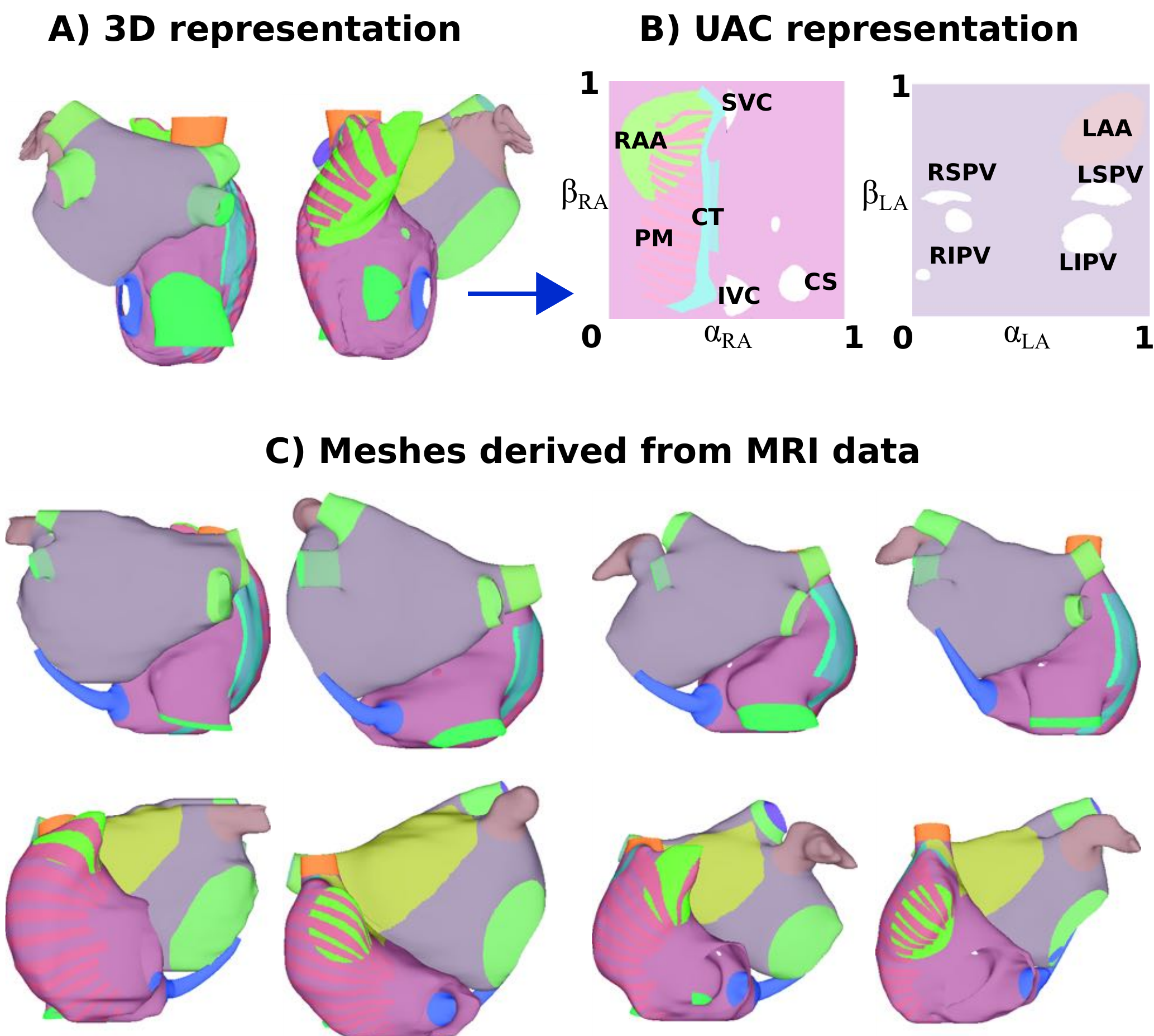}
\caption{\textbf{Patient specific geometries:} (A) The original bilayer model displayed in posteroanterior and anteroposterior view. RA endocardial structures are displayed as epicardial, for visualisation purposes. (B) 2D UAC representation with regions labelled for the RA (left) and LA (right). (C) Meshes derived from MRI data for four patients.}
\label{fig:PatientGeometries}

\end{figure}

 Fig.~\ref{fig:PatientGeometries} shows atrial structures for the reference bilayer geometry (Fig.~\ref{fig:PatientGeometries} A), mapped to four patient specific geometries (Fig.~\ref{fig:PatientGeometries} C). Visually the atrial structures are located in similar locations across meshes. This demonstrates the utility of UACs for constructing meshes from patient imaging data with atrial structures, allowing the incorporation of regional heterogeneity in the model. The atrial regions are also shown in 2D for the LA and the RA (Fig.~\ref{fig:PatientGeometries} B), demonstrating the intermediate step of the mapping.

 \subsection{Fibre mapping and comparison to registration techniques}
\begin{figure}[!t]
\centering
\includegraphics[width=0.45\textwidth]{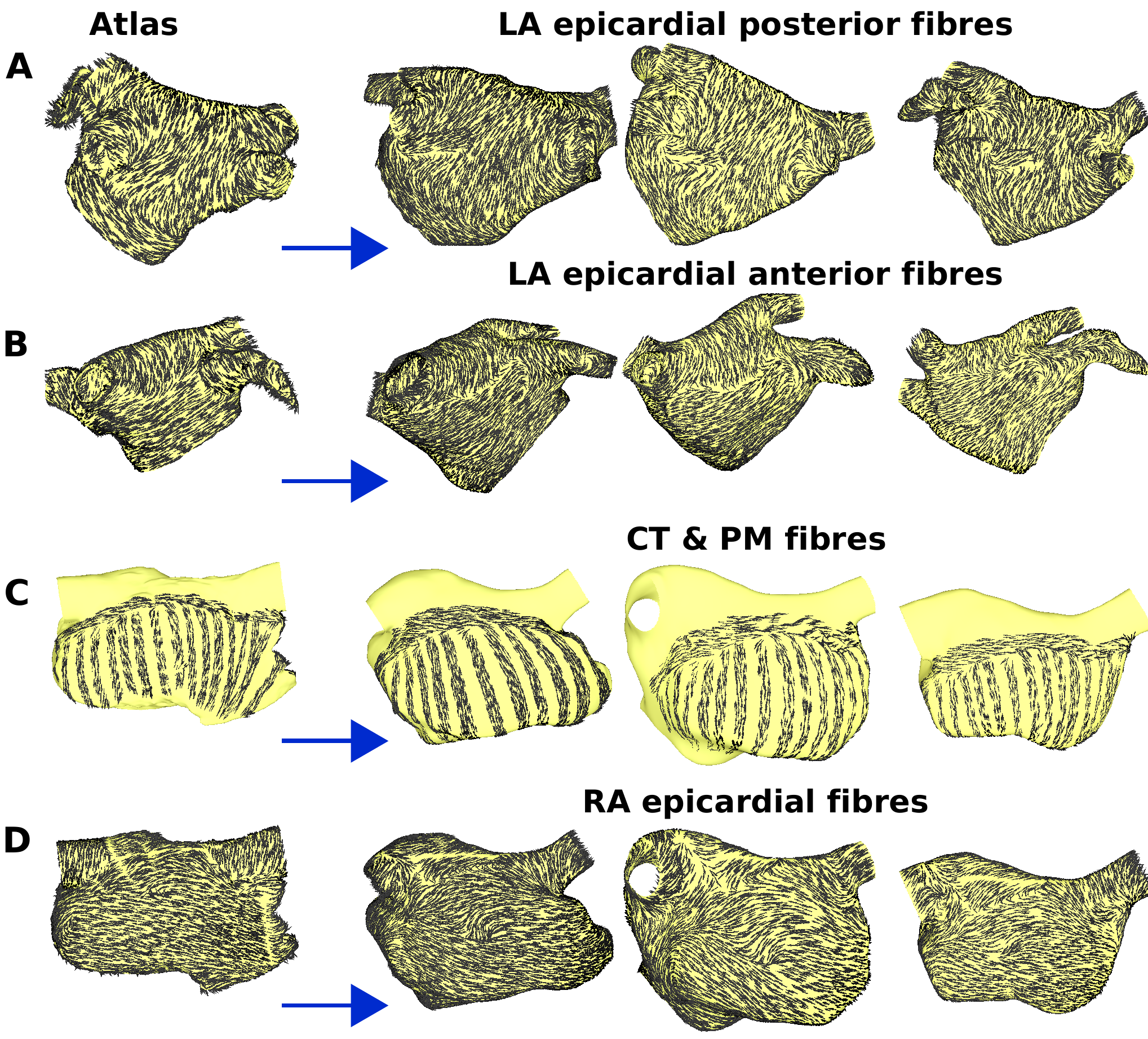}
\caption{\textbf{Fibre mapping:} The original bilayer geometry is shown on the left column, with fibres for the: (A) LA epicardium, posteroanterior view; (B) LA epicardium, anteroposterior view; (C) CT and PM; (D) RA epicardial fibres.}
\label{fig:fibremaps}

\end{figure}

 Fig.~\ref{fig:fibremaps} shows fibre directions for the original bilayer model, as well as fibre directions mapped to different patient geometries. The mapping demonstrates the use of UACs for constructing patient specific meshes with fibre direction.

\begin{figure}[!t]
\centering
\includegraphics[width=0.45\textwidth]{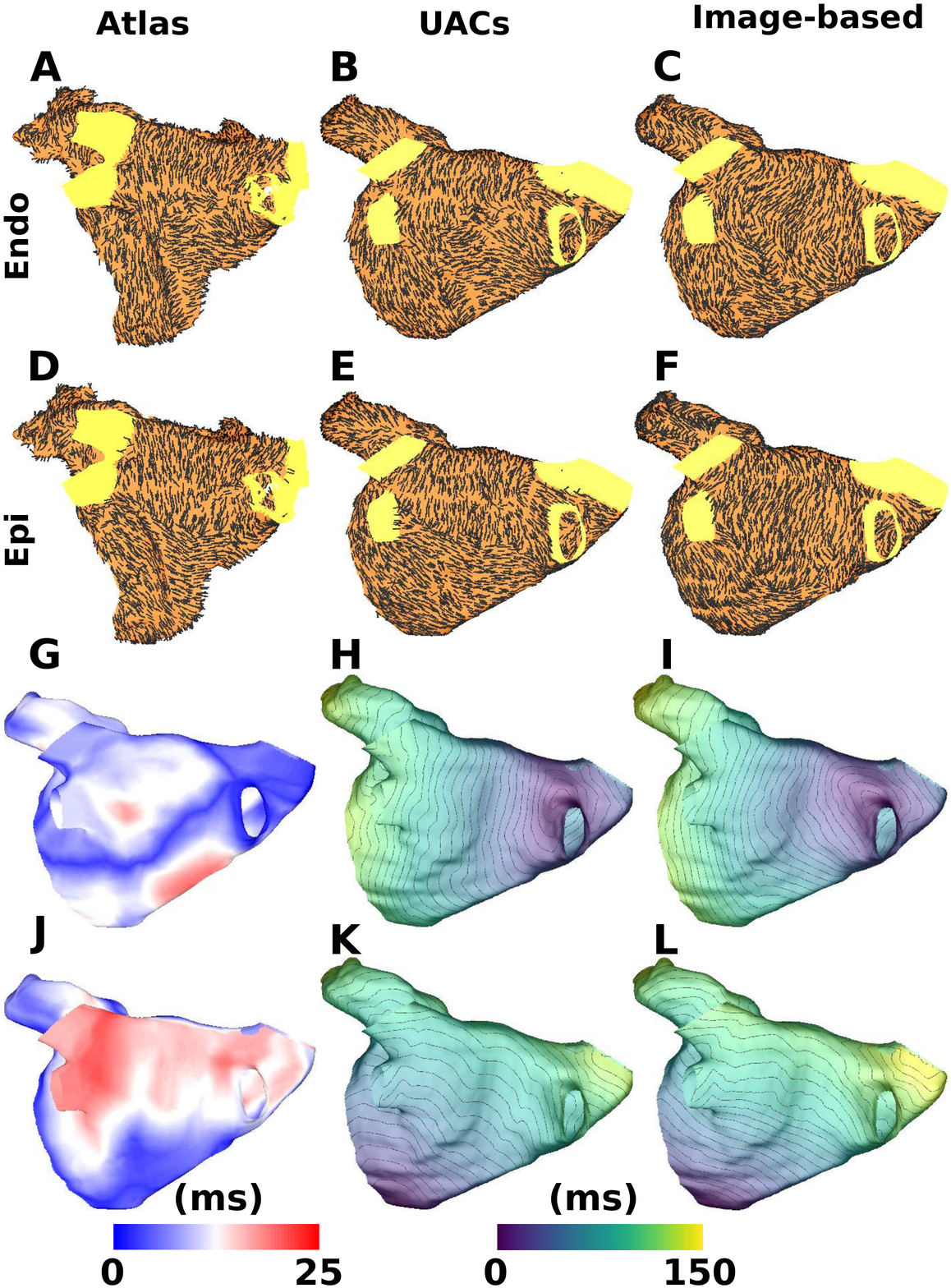}
\caption{\textbf{Comparison of fibre mapping techniques:} Endocardial (A-C) and epicardial (D-F) fibre directions for a human atrial atlas (A$\&$D) mapped to a different patient geometry using either UACs (B$\&$E) or an image based method (C$\&$F). Differences in activation time for the endocardium and epicardium with pacing from the roof or MV respectively are shown in (G) and (J). Activation time maps corresponding to (B, C) are shown in (H, I) for roof pacing, and for (E, F) in (K, L) for MV pacing.  
}
\label{fig:fibvalidation}
\end{figure}

The technique developed in this paper for mapping of fibre directions using UACs was compared to fibres assigned using a previously published image based mapping method \cite{mcdowell2012methodology}. This image based method uses a manual land-marking process followed by a 3D thin plate spline transformation \cite{turk2005shape} and a large deformation diffeomorphic metric mapping \cite{beg2005computing} to transform an atlas geometry to a patient specific geometry. 
The optimal deformation field calculated by this image based method is then applied to a vector field of atlas fibres to obtain mapped patient specific fibre orientations.
The patient specific geometry considered here is from Zahid et al. \cite{zahid2016patient} for which fibres were assigned using the image based mapping method with the atlas geometry of Krueger et al. \cite{krueger2011modeling}. 
To compare our mapping methodology to that of McDowell et al.  \cite{mcdowell2012methodology}, fibres were mapped using both methods from a human atrial atlas \cite{krueger2011modeling}, where only the endocardial and epicardial surface fibres were considered (Fig~\ref{fig:fibvalidation} A and D). 

There is spatial variation in the degree of correspondence between the methods, with a degree of agreement on the lower posterior wall and roof, and more differences on the mid posterior wall (compare Fig~\ref{fig:fibvalidation} B to C for endocardial fibres, as well as E to F for epicardial fibres). To spatially quantify the effects of these differences in cardiac model simulations, activation was simulated with stimulation from two different pacing locations for the four different fibre arrangements (Fig~\ref{fig:fibvalidation} B, C, E and F), with conductivities of 0.4 S/m in the longitudinal direction and 0.1 S/m in the transverse direction.
Differences between endocardial and epicardial activation for the same method were small (mean total activation time difference: 4.1 $\pm$ 3.1 ms; mean point-wise activation time difference: 3.7 $\pm$ 2.2 ms). 
Differences between the two methods were larger (mean total activation time difference: 12.2 $\pm$ 5.3 ms; mean point-wise activation time difference: 7.2 $\pm$ 2.1 ms). 
An example map of these activation differences for the endocardium is shown in Fig~\ref{fig:fibvalidation} G, in which differences are seen on the posterior wall below the RIPV due to differences in the fibres between Fig~\ref{fig:fibvalidation} B and C in this region. 
Fig~\ref{fig:fibvalidation} H and I show the corresponding activation time maps for roof pacing for the endocardial UAC fibres in B, and the image-based fibres in C, respectively. 
Fig~\ref{fig:fibvalidation} J shows that there are large differences in activation time on the roof of the epicardial LA when pacing from the MV. 
Comparing the activation time maps in Fig~\ref{fig:fibvalidation} K and L shows that the isochrones on the roof are similarly spaced, and the differences seen in Fig~\ref{fig:fibvalidation} J are due to large differences in the fibre directions on the lower posterior wall between Fig~\ref{fig:fibvalidation} E and F, rather than differences in fibres on the atrial roof.

\subsection{Sensitivity to choice of markers} \label{sec:sensitivity}

We investigated the sensitivity of the UAC mapping to assignment of the UAC landmark points. 
The left atrial coordinates depend on the points selected at the LSPV-LA and RSPV-LA junctions.
This is because this path is used for the roof Dirichlet boundary condition for $\beta_{LA}$. To a lesser extent, this choice also affects $\alpha_{LA}$ since the roof line determines the location of the isoline used for rescaling. 
The choice of the third LA point, which is just anterior of the FO affects the location of the septal boundary condition for $\alpha_{LA}$. For equivalent reasons, the right atrial coordinates depend on the points selected at the SVC-RA and IVC-RA junctions. 
We consider the MV and TV ring positions as fixed, since these are assigned during the segmentation and mesh modification step of the pipeline. 

\begin{figure}[!t]
\centering
\includegraphics[width=0.45\textwidth]{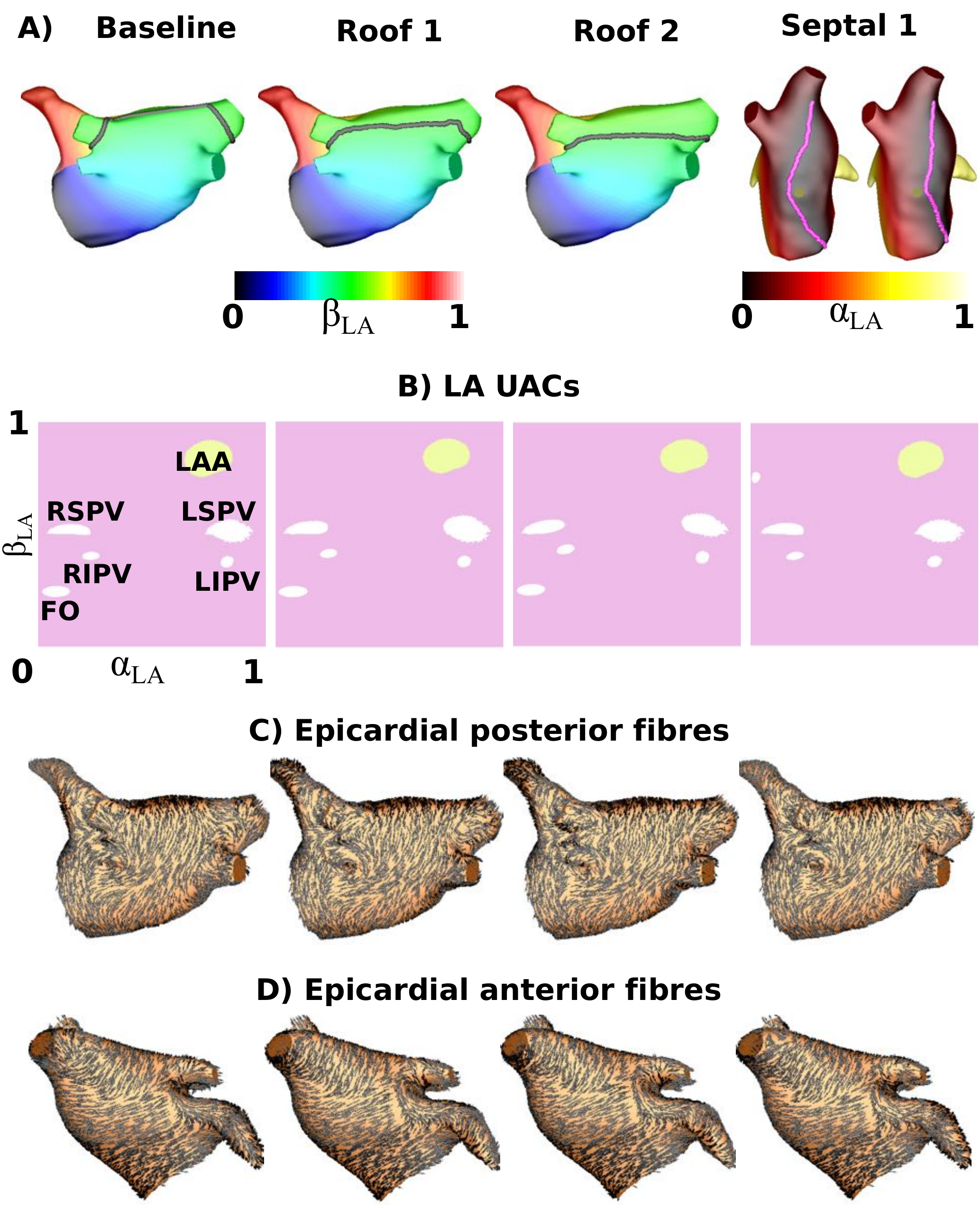}
\caption{\textbf{Point sensitivity.} (A) Boundary nodes used for the Laplace solves for the baseline case, a case with the LSPV and RSPV markers moved to the middle of the veins (roof 1), a case with the markers at the base of the veins (roof 2), and a case with the septal marker moved to be posterior of the FO. These changes visibly change the boundary condition locations. (B) LA UACs. (C) Epicardial posterior fibres. (D) Epicardial anterior fibres.} 
\label{fig:PS}

\end{figure}

The sensitivity of the mapping to the five points selected for the UAC assignment was determined. An example is shown in Fig ~\ref{fig:PS} in which the LSPV and RSPV points were modified to be at the middle and base of the LA-PV junctions. As a final example, the septal boundary marker was moved to be posterior of the FO.
Modifying the points as such resulted in the following angle errors: 
middle roof assignment (median angle error: 6.0$^\circ$),
base roof assignment (median angle error: 8.1$^\circ$),
posterior FO (median angle error: 7.2$^\circ$).

 \subsection{Visualisation}
\begin{figure}[!t]
\centering
\includegraphics[width=0.45\textwidth]{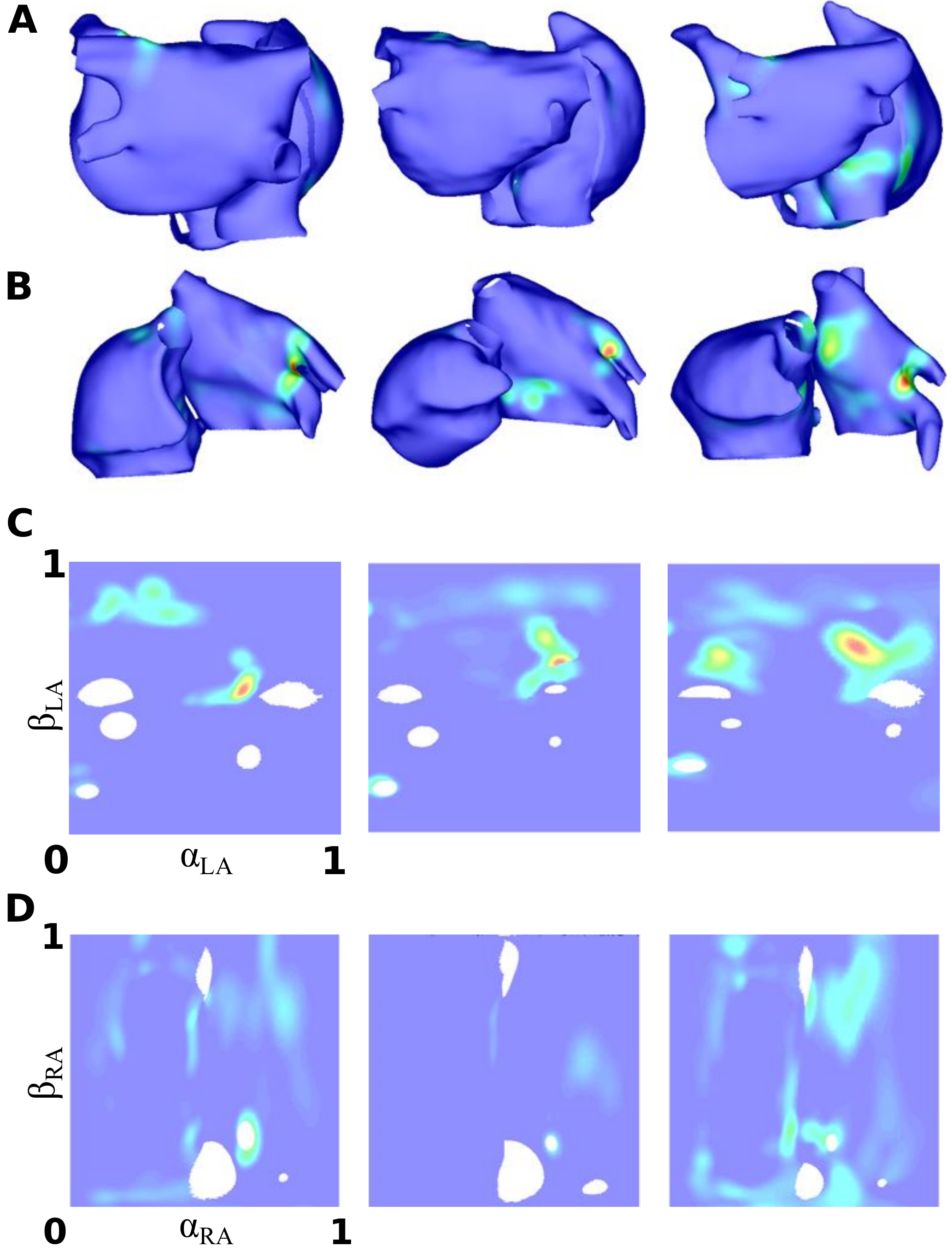}
\caption{\textbf{Phase singularity plots visualised in 2D:} Normalised phase singularity density maps for three patients in: (A) Posteroanterior view; (B) Anteroposterior view; (C) LA 2D representation; (D) RA 2D representation. Orientation as per Fig.~\ref{fig:PatientGeometries}.}
\label{fig:PSmaps}

\end{figure}

 Fig~\ref{fig:PSmaps} shows phase singularity distributions displayed on different atrial geometries, as well as in two-dimensional UACs. These distributions vary across geometries. The use of UACs for two-dimensional visualisation allows each of the LA and RA to be displayed in a single map (LA shown in Fig~\ref{fig:PSmaps} C; RA in Fig~\ref{fig:PSmaps} D), so that multiple views are not necessary. Converting each map to the same 2D coordinate system also allows correlations to be calculated between different geometries. These maps are for the LA and RA epicardial surfaces. Other atrial structures can be assessed separately, as shown in Fig.~\ref{fig:PatientGeometries} B.

 \section{Discussion}
In this paper, we present a universal atrial coordinate system that can be used to map scalar and vector data between any two geometries; to map multiple data sets recorded from an individual and also registering datasets recorded across a population; to visualise atrial data in two dimensions; and to construct patient specific bilayer computational models, with electrophysiological heterogeneity and fibre direction. Our methodology requires the manual selection of only five points to create a patient specific mesh with regional heterogeneity and fibre direction. 

 Our methodology allows 2D visualisation on a square geometry for both the LA and RA, and as such is similar to the visualisation of  Karim et al. \cite{karim2014surface} and Williams et al. \cite{williams2017standardized}. This 2D representation (see Figs~\ref{fig:PatientGeometries} $\&$ ~\ref{fig:PSmaps}) in a standard coordinate system allows easy regional comparison across multiple patients, calculation of correlation between measures, and provides a format compatible with CNN analysis. 
 For clinical utility, we used the same LA and RA orientation as is used for 2D visualisation of biatrial basket catheter data \cite{narayan2012computational}. 
 
 Point transfer using UAC was shown to have a high accuracy (see Table~\ref{tab:pointtransfer}), with all mean errors much smaller than the average edge length. 
 UAC scalar mapping performed similarly to a standard affine registration technique using MIRTK when mapping from a low resolution electroanatomic mapping mesh to a higher resolution MRI geometry (see Fig~\ref{fig:Carto}). 
 However, the UAC method is independent of the initial orientations and coordinates of the two meshes, making it more robust to mesh location initialisation. 
 In addition, UAC mapping may offer better performance than affine registration for cases in which the two atrial geometries have very different morphologies. 
 
 Our atrial pipeline uses UACs to assign both atrial structures for regional heterogeneity and fibre directions to the mesh. As such, a single technique, with the manual selection of only five points (see Fig~\ref{fig:UACpoints}) can generate a patient specific mesh (see Fig~\ref{fig:PatientGeometries}) with fibre direction (see Fig~\ref{fig:fibremaps}).
Alessandrini et al. use a rule-based technique with a set of 13 anatomical landmark points to assign fibres and regions to an LA model, for which the location of BB and the FO were assigned manually \cite{alessandrini2018computational, wachter2015mesh}. 
 Many modelling studies only include the LA or else do not include atrial structures explicitly. 
 
 Our current pipeline for constructing patient specific meshes uses a previously published biatrial bilayer model \cite{labarthe2014bilayer} as the reference geometry, to inform locations of atrial regions and fibre directions. As such, all patient specific models have specific geometries from patient MRI data, but the morphology of BB, the PM, CT, the line of block and the SA node are similar to the original bilayer geometry. These structures cannot easily be determined from current MRI resolution, necessitating this mapping approach. 
 The CS was included as a cylindrical structure with a single connection point to the LA; however, the model could easily be modified to include multiple connection points. 
 
 Similarly, fibres are currently mapped from the original bilayer model for which fibres were included based on histological descriptions \cite{ho1999anatomy}, combined with a rule-based approach \cite{labarthe2012semi}.
 The fibres assigned using our UAC methodology (see Fig~\ref{fig:fibremaps}) are different to those using a image based approach (see Fig~\ref{fig:fibvalidation}); however, it is difficult to determine which distribution is more appropriate. 
 The methodology introduced in our paper is general, and can easily use a different vector map as an input. For example, Zhao et al. measure high-resolution fibre data in sheep atria using histology \cite{zhao2012image}, and the recent study of Pashakhanloo et al. provides high resolution DT-MRI data, which could be used as an alternative fibre field input for the mapping \cite{pashakhanloo2016myofiber}. 
 
During the development of the methodology, multiple choices of coordinates were tried and compared. Our initial motivation for the choice of coordinates was that they should be as close to orthogonal as possible and use clear atrial landmarks. We initially tried a LA lateral-septal coordinate with boundary conditions of 0 at the LAA and 1 at the FO. However, close to the FO,  the field was radial so this region of the mesh was not well defined by the coordinate system. Using the LAA as a boundary did not create such a problem since it is larger; however, the size and shape of the LAA-LA boundary varies between meshes, and instead using a linear boundary was found to be optimal. An alternative choice for the origin of the LA coordinate system is the centre of the line connecting the median points of the left and right PVs, as suggested by Pashakhanloo et al. \cite{pashakhanloo2016myofiber}.  

Five points were manually selected for UAC assignment; four of these points were on the veins (see Fig~\ref{fig:UACpoints}). Previous studies have found that landmarks at the PV-LA junction are the most accurate choice for LA registration \cite{fahmy2007intracardiac, ali2015automated}, which offers further justification for our choice of markers. The UAC system requires the selection of points on the superior veins; as such, the coordinate system is still applicable for patients with three or five veins as long as one vein is chosen to represent the LSPV and another the RSPV.  

\subsection{Limitations} 
This UAC system could be extended to three dimensions by the inclusion of a transmural coordinate, by solving a Laplace equation with 0 on the endocardial surface and 1 on the epicardial surface as in \cite{bishop2016three}, to incorporate transmural fibre assignment. In addition, the technique presented here for mapping vector data could be adapted to instead use deformation gradients as for the UVC system \cite{bayer2018universal}. The degree of distortion in the UAC space gives an indication of the morphology of the atrial anatomy, and could be used as a geometry metric \cite{bisbal2013left}. 
The inclusion of BB and the CS in the biatrial bilayer model required additional methodology as these structures are not well represented in the UAC system. The degree of reproducibility in region assignment could be improved for the PVs by using an automated technique and the system could be integrated into a single tool for clinical utility; for example, following \cite{razeghi2017platform}.

\subsection{Conclusions} 
We have developed a novel coordinate system that has multiple applications, including mapping scalar and vector data between any two atrial geometries, visualising data in 2D and constructing patient specific atrial geometries. We demonstrated that a patient specific bilayer biatrial geometry with repolarisation heterogeneity and fibre direction can be constructed with the manual selection of just five points. The technique was validated against standard registration techniques for both scalar and vector mapping. 

\section*{Acknowledgements}
We would like to thank Prof Elliot McVeigh and Dr Farhad Pashakhanloo for helpful discussions on the coordinate system. 

\section*{Funding sources}
CHR acknowledges a Lefoulon-Delalande Foundation fellowship administered by the Institute of France. 
This study received financial support from the French Government as part of the Investments of the Future program managed by the National Research Agency (ANR), grant reference ANR-10-IAHU-04.
We acknowledge PRACE for awarding us access to ARCHER UK National Supercomputing Service. 
SAN acknowledges support from the UK Engineering and Physical Sciences Research Council (EP/M012492/1, NS/A000049/1 and EP/P01268X/1), the British Heart Foundation (PG/15/91/31812, PG/13/37/30280) and Kings Health Partners London National Institute for Health Research (NIHR) Biomedical Research Centre. This work was supported by the Wellcome/EPSRC Centre for Medical Engineering [WT 203148/Z/16/Z].

\bibliographystyle{ieeetr}

\bibliography{library.bib}



\section{Supplementary Material}

\subsection{Assigning atrial regions} \label{sec:assignPVsApp}
\begin{figure}[htbp]
\centering
\includegraphics[width=0.45\textwidth]{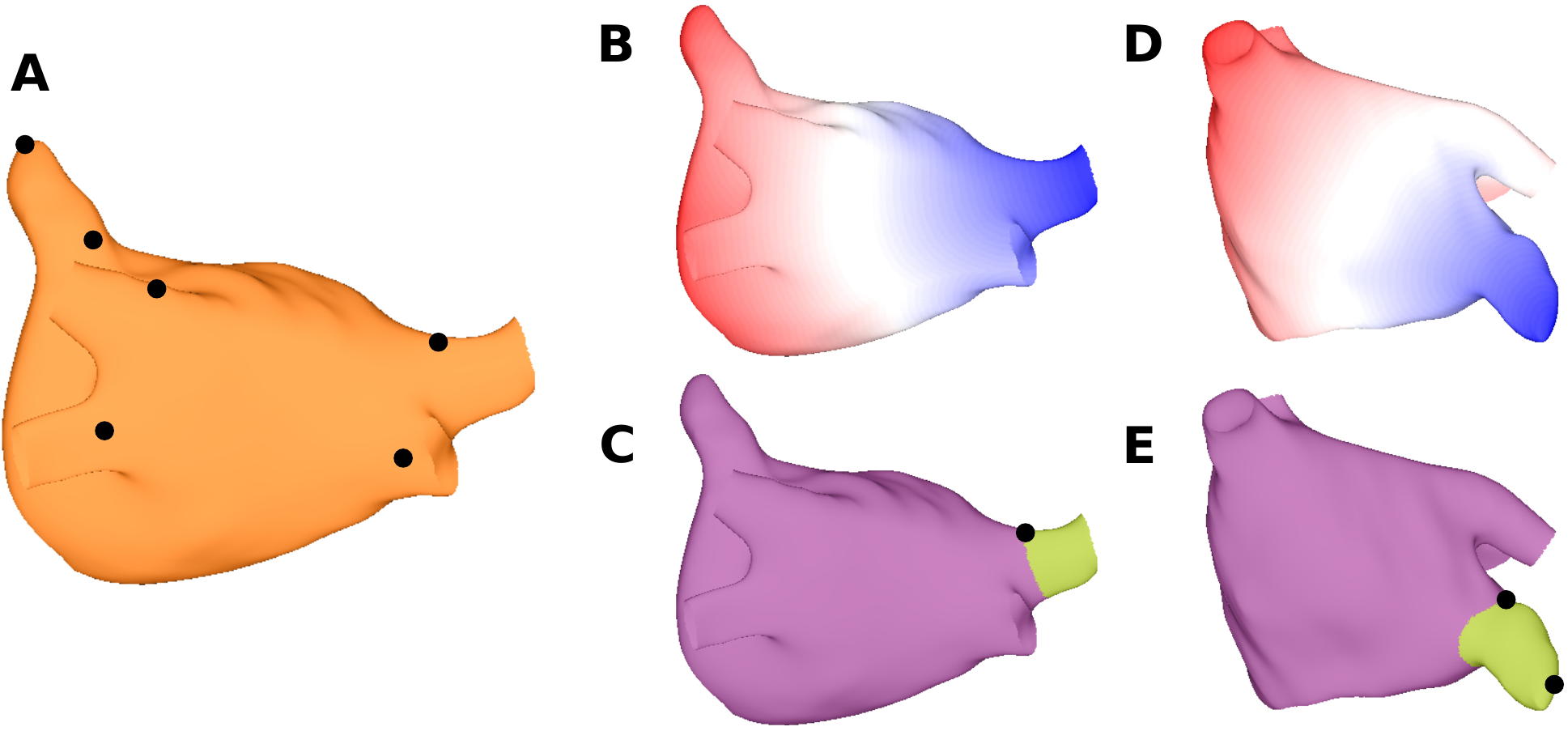}
\caption{\textbf{Assignment of veins and appendages:} (A) Example points selected for assignment of each region. (B) Geodesic map for the RSPV. (C) Points with geodesic distance less than the selected point in black are assigned to the RSPV region. (D) Geodesic map for the LAA. (E) Points with geodesic distance less than the selected point at the LAA-LA junction (shown in black) are assigned to the LAA region. 
}
\label{fig:regions}
\end{figure}

In this study, regions were labelled by selecting points at the boundary of the structure and the atrial body and using a geodesic distance criterion, as follows. 
For the LA, one point was selected per PV and two points for the appendage. For the RA, a point was selected for each of the SVC, IVC and CS, and two points were selected for the appendage. These points are shown in Fig~\ref{fig:regions}, and details for their selection are as follows. 

For each of the PVs, the SVC, and the IVC, a point was selected at the boundary of the structure and the atrial body (see Fig~\ref{fig:regions} A). 
For the left and right atrial appendages (LAA and RAA), one point was selected on the tip of the appendage, and a second point was selected on the boundary of the appendage and atrial body (see Fig~\ref{fig:regions} E). 

The selected points were then used to define atrial structures using the following methodology.  
First of all, the boundary edges of the left and right atrial meshes were identified, and separated into lists of connected edges (representing the boundaries of the MV and TV; each of the PVs; the SVC, IVC and CS). 
For each of the PVs, the SVC and the IVC, geodesic paths were calculated from the nodes in the boundary edge list to the rest of the mesh. Then for each node of the mesh, the minimum distance to the set of nodes in the boundary edge list was calculated (see Fig~\ref{fig:regions}B). Regions were then defined as the set of nodes closer to the nodes in the boundary edge list for that region than the selected point (see Fig~\ref{fig:regions}C).

A similar technique was used to define the left and right atrial appendages. Geodesic distances were calculated from the point at the tip of the appendage to the rest of the mesh (see Fig~\ref{fig:regions}D), and nodes closer to the tip than the point selected at the boundary of the appendage and atrial body, were defined as appendage (see Fig~\ref{fig:regions}E). 
Geodesic paths were calculated using a marching cubes algorithm \cite{peyre2009toolbox, roney2015technique}. 

\subsubsection{Incorporating additional structures - Bachmann's bundle and the coronary sinus}\label{sec:BBCS}
Bachmann's bundle and the coronary sinus require additional steps for their construction because these structures are not fully determined by the UAC system. Part of the CS structure was assigned to the RA mesh that was generated from MRI data (see Section~\ref{sec:assignPVsApp}) and expressed in UACs, and this was then extended to a cylindrical structure. This was implemented by identifying the rim of boundary nodes of the CS structure on the MRI mesh, projecting these to a plane, and calculating a center line trajectory for the CS cylinder. This trajectory was constructed by selecting a target end point on the LA mesh, finding a geodesic path from this point to the point on the LA closest to the CS boundary nodes and then projecting this path off of the mesh. A spline was then constructed to join the distal nodes of the path to a trajectory from the centre point of the boundary rim that is tangential to the normal of the boundary plane. The resulting path is then in the tangential direction to the initial CS structure at the start of the cylinder, and follows the morphology of the LA mesh. The boundary nodes were then marched forward by shifting the boundary plane along the tangent to the centre line trajectory, for both this arrangement, and for nodes at the mid-points, in order to create triangular elements. To introduce a degree of taper, the diameter of the bounding circle of the nodes was decreased along the length of the cylinder. This is demonstrated in Supplementary Fig~\ref{fig:CS_BB} A.

 \begin{figure}[htbp]
\centering
\includegraphics[width=0.45\textwidth]{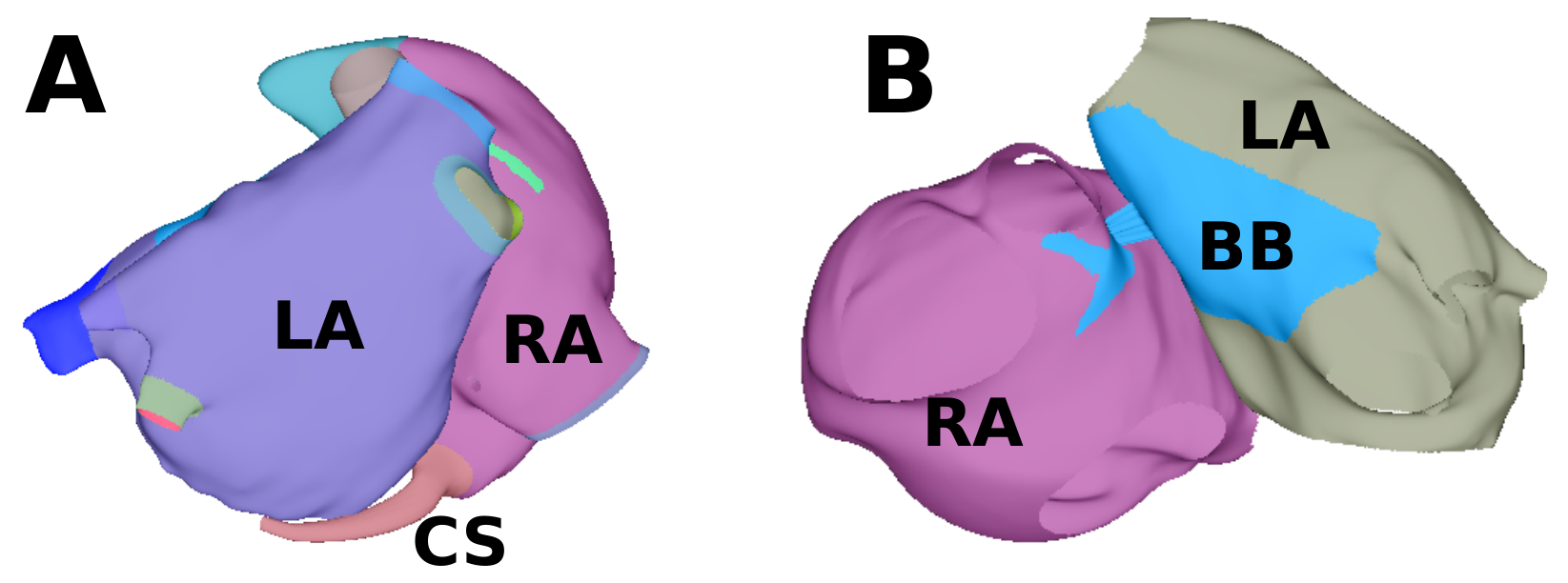}
\caption{\textbf{Including the coronary sinus and Bachmann's bundle in the model:} (A) The inclusion of the CS. (B) The inclusion of BB.}
\label{fig:CS_BB}
\end{figure}

LA and RA components of BB were initially assigned using scalar mapping as for the other structures. It is not possible to assign the component of BB between the LA and RA using UACs because this region is not close enough to either mesh. Instead, the LA and RA components of BB were joined to each other using an automatic mesh construction technique. An equal number of connected nodes along the boundary of each component were selected and then new nodes were distributed at an equal spacing along the vector joining equivalent nodes on each component, and finally nodes were added at mid-points to construct triangular elements. 
This is demonstrated in Supplementary Fig~\ref{fig:CS_BB} B.

\subsubsection{Connections}\label{sec:connections}
To construct a bilayer model, discrete resistance line connections were added between each vertex of the LA endocardial mesh and the closest LA epicardial vertex. Connections between the sinus node (SAN) and RA were assigned to the model by mapping the original mesh SAN line connection vertices to the new mesh using UACs. The LA and RA FO rims were electrically connected by joining the closest LA and RA nodes using line connections. In addition, the half of the distal CS rim closest to the LA body was joined to the closest LA nodes using line connections.

\end{document}